\documentclass[letterpaper,onecolumn]{article}
\usepackage{times}
\usepackage{graphicx}
\usepackage{amsmath}
\usepackage{amssymb}
\usepackage[top=1.50cm,left=1.50cm,right=1.50cm,bottom=1.50cm]{geometry}

\begin{document}

% Don't want date printed
\date{}

% Make title bold and 14 pt font (Latex default is non-bold, 16pt)
\title{\Large\bf Spectral Efficiency of Multiple Access Fading Channels with Adaptive Interference Cancellation}

%\normalsize
%\normalsize
% ICCV PAPER SUBMISSIONS MUST BE ANONYMOUS. DO NOT PUT
% YOUR NAME HERE.  PLEASE PUT YOUR ICCV PAPER NUMBER IN
% THIS FIELD IF YOU KNOW IT.
\author{\small  Indu L. Shakya, Falah H. Ali}
\maketitle \baselineskip=2.0 \baselineskip

%\section*{\ Abstract}
\noindent\textbf{Abstract}

Reliable estimation of users' channels and data in rapidly time
varying fading environments is a very challenging task of multiuser
detection (MUD) techniques that promise impressive capacity gains
for interference limited systems such as non-orthogonal CDMA and
spatial multiplexing MIMO based LTE. This paper analyzes relative
channel estimation error performances of conventional single user
and multiuser receivers for an uplink of DS-CDMA and shows their
impact on output signal to interference and noise ratio (SINR)
performances. Mean squared error (MSE) of channel estimation and
achievable spectral efficiencies of these receivers obtained from
the output SINR calculations are then compared with that achieved
with new adaptive interference canceling receivers. It is shown that
the adaptive receivers using successive (SIC) and parallel
interference cancellation (PIC) methods offer much improved channel
estimation and SINR performances, and hence significant increase in
achievable sum date rates.

\noindent \textbf{\large\bf Keywords}: \textbf{Multiuser Detection,
Interference Cancellation, Channel Estimation}\\

\noindent Corresponding Author: \textbf {Dr. Indu L. Shakya},\\
Communications Research Group,\\
School of Engineering and Informatics\\
University of Sussex, Brighton, UK, BN1 9QT\\
Email: i.l.shakya@sussex.ac.uk, Tel: (44)-1273-678445

\pagebreak

\section{Introduction}
% no \PARstart
Study of achievable spectral efficiency of realistic interference
limited systems such as uplink of CDMA and spatial multiplexing
based MIMO systems using different detection techniques is very
important from practical points of view. These systems have many
similar characteristics: for example, decorrelators or zero forcing
detectors used for the both systems are based on the same principle
of nulling of interference subspaces and interference cancellation
stage is usually added to further enhance the detection performance.
Therefore, the techniques and analyses designed for CDMA uplink, are
generally applicable to both these systems. There exists many
studies on efficient detection techniques for CDMA and MIMO systems,
e.g. \cite{Verdu}, \cite{PER_SIC}, \cite{MMSE_SIC}, \cite{MIMO_LTE}.
It has been noted in \cite{varanasi}, an efficient interference
cancellation technique combined with multiuser detection is very
important to achieve the channel capacity.

The MUD schemes using interference cancellation techniques such as
decorrelating decision feedback detectors and V-BLAST usually suffer
from error propagation at each stage and requires very accurate
channel estimation to maximize the achievable performance
\cite{vblast_est}. To address this problem, the authors have
proposed new blind adaptive SIC \cite{shakya_sic2007} and PIC
\cite{shakya_pic} techniques that exploits the interesting amplitude
restoring property of the well known constant modulus algorithm
(CMA) to minimize the error propagation. Referred to here as BA-SIC
and BA-PIC, these schemes have shown to achieve much improved
interference estimation and cancellation performances to minimize
the error propagation.

Analyzing the spectral efficiency of different receivers under
realistic mobile Rayleigh fading channels with estimation errors is
very interesting and challenging problem. Asymptotic spectral
efficiency analysis of CDMA with different multiuser receivers using
random spreading sequences in simple additive white Gaussian noise
(AWGN) environment has been shown in \cite{spec_verdu}. The spectral
efficiency analysis of SIC receivers in Rayleigh fading with perfect
channel estimation is carried out in \cite{djonin_sic}. In the case
of PIC, signal to interference and noise ratio (SINR) and system
capacity analysis of a linear PIC is carried out in \cite{brown_pic}
for simple AWGN channel environment.

These work have either assumed static channel or fading channels
with perfect channel estimation. In practice, users' channel are
often time varying and with high Doppler rates. In such environment,
estimation of desired users' channels/data while fully suppressing
interfering users' signals is very challenging task. In this paper,
we investigate channel/interference estimation and cancellation
performance of conventional single user and multiuser receivers in
comparison with the new adaptive IC receivers \cite{shakya_sic2007},
\cite{shakya_pic} in mobile Rayleigh fading channels. We also derive
their respective achievable spectral efficiencies based on output
SINR calculations to show the significant gains of the new adaptive
approaches.

This paper is organized as follows. A general system model is
described Section II. The estimation approaches of the adaptive
schemes are described in Section III. The sum rate or spectral
efficiency analysis of the schemes to be compared is carried out in
Section IV. The performance study in terms of mean squared error of
channel estimation and achievable spectral efficiency is provided in
Section V. Finally the paper is concluded in Section VI.

\section{System Model} \label{sys_model}
An uplink of synchronous DS-CDMA system of $K$ users under Rayleigh
flat fading and AWGN channel conditions is considered here. The
received composite signal $\textbf{r}=[r_1,r_2,..,r_N]^{T}$ can be
written as:
\begin{equation}
\label{eqn:kusers_rayleigh}
 \textbf{r}(m){=}\sum_{k=1}^K \beta_k(m) b_k(m) \textbf{c}_k(m)+\textbf{v}(m),
\end{equation}
where $b_k$ is the data signal of period $T_b$ and satisfying
condition $E\{b_{k}^2\} \leq P_k$ where $P_k$ is the $k^{th}$ user's
signal power. The spreading sequence is denoted as $\textbf{c}_k$
with antipodal chips with period $T_c$ and normalized power over a
symbol period equal to unity $\int_{0}^{T_{b}}c_k(t)^2dt=1$. The
spreading factor is $N=T_b/T_c$ and $\textbf{v}$ is the AWGN with
two sided power spectral density $N_0/2$, $\beta_k(n)=g_k(n)
e^{-j\phi_k(n)}$ is sample of time varying complex Rayleigh flat
fading channel with zero mean and unit variance and consisting of
amplitude $g_k$ and phase $\phi_k$ components, respectively. Note
that flat fading channel model is used here for the simplicity; the
frequency selective channels can also be considered in proposed
system model by using the method in \cite{djonin_sic} by modeling
each multipath signal as independently faded and uncorrelated with
the desired path. Where it is also shown that spectral efficiency of
a SIC can be higher than that in flat fading case and hence analysis
of our schemes can also be extended to multipath channels.

Using the generic system model above, main ideas behind different
receiver techniques and the new adaptive methods are briefly
described next. For all receivers, it is assumed that coherent phase
reference with perfect knowledge of phases of all users' channels
are available and without loss of generality $k^{th}$ user is the
desired user. Note that channel phase estimation can be achieved at
the receiver by using the known pilot sequences inserted in users'
data streams and is widely used in existing wireless systems.

\noindent \textbf{Conventional Receivers/Matched Filters (MF):} The
MF detectors obtain data and channel estimates using the despreader
output $z^{MF}_k$, also given by
\begin{equation}
z^{MF}_k=\int_{0}^{T_b}\textbf{r}_k \textbf{c}^{T}_k= g_k
b_k+\sum_{j=1, j \neq k}^{K} \rho_{kj}g_j b_j+v_k
\end{equation}
where the three terms are the desired signal, the sum of interfering
users' data signals, and the AWGN, respectively and $\rho_{ki}$ is
the magnitude of cross-correlation between $k^{th}$ and $i^{th}$
users' spreading sequences. As can be noted here that for the MF,
the channel and data estimate are estimated jointly and is simply
taken as
\begin{equation}
\hat{g_k} b_k= z_{MF}.
\end {equation}

This leads to the SINR $\Gamma$, expression of the receiver as:
\begin{equation}
\Gamma^{MF}_k=\frac{E(z^{MF}_k)^2}{var(z^{MF}_k)}=\frac{E\big\{g_k
b_k \big\}^2}{E\big\{\sum_{j\neq k}^{K}\rho_{kj}g_j
b_j\big\}^2+N_{0}}
\end {equation}
where $E\{.\}$ and $var\{.\}$ denote the expectation and variance,
respectively of a random variable.

\noindent \textbf{Conventional PIC:} The PIC receivers employ
parallel IC processes one for each user and iterative detection
using multiple i. e. $\geq 2$ stages to refine the users' data
estimates. The first stage of a PIC is usually the bank of matched
filters hence it's performance at this stage is the same as that of
the conventional MF receivers. The decision variable at $l^{th},
l\geq 0$ stage $z^{PIC(l)}_k$ is obtained by cancelling from
$\textbf{r}$ the summation of the other users' signals as follows:
\begin{equation}
z^{PIC(l)}_k=\int_{0}^{Tb}\Big\{\textbf{r}-\sum_{j=1,j \neq
k}^{K}z^{PIC(l-1)}_j\textbf{c}_{j}\Big\}^{T}\textbf{c}_{k}; \forall
k.
\end{equation}

The data and channel estimate at $l^{th}$ stage is taken as
\begin{equation}
\hat{g_k}^{PIC(l)} b_k= z^{PIC(l)}_k.
\end{equation}
As can be seen conventional PIC reduces to MF receiver at $l=0$. The
SINR for a PIC is dependent upon the estimation accuracy of the MF
detectors used for each user i. e. by the mean square error (MSE)
which will be give later in equation (\ref{eqn:min_mse}). Using the
MSE measure, the SINR at $l$ stage can also be obtained as:
\begin{equation}
\Gamma^{PIC(l)}_k=\frac{E\big(g_k b_k
\big)^2}{E\big\{\sum_{j=1}^{K}(\hat{g}^{PIC(l)}_j-g_j)^2 \rho_{kj}
b_j\big\}^2+N_{0}}.
\end {equation}

\noindent \textbf{Conventional SIC:} The SIC technique operates on
the principles of ordering of users' signals based on their received
power estimates for performing decision on the strongest and then
subtracting the estimated signal from the total remaining received
signal. The decision variable at each stage $k, 1\leq k \leq K$, is
obtained as
\begin{equation}
z^{SIC}_k=\max_{1 \leq u \leq
K}\Big\{{\int_{0}^{Tb}\textbf{r}_{k}^{T}\textbf{c}_u}\Big\}.
\end{equation}

The data and channel estimate is taken as
\begin{equation}
\hat{g_k} b_k= z^{SIC}_k.
\end{equation}

The signal $z^{SIC}_k$ is respread and subtracted from the remaining
received signal to remove its interference as follows:
\begin{equation}
\label{sic} \textbf{r}_{k+1}=\textbf{r}_{k}-z^{SIC}_k \textbf{c}_k.
\end{equation}

These processes are carried out until all users' data are detected.
The SINR for the case of SIC is variable for each user as the
earlier detected user sees more interference while the least
detected user may not see any interference. Although this assumption
is inaccurate due to the imperfect estimation of each users' signal,
we adopt this simplified approach to just reveal the relative SINR
gain of the SIC technique. The SINR for a user using the SIC
technique can be shown as:
\begin{equation}
\Gamma^{SIC}_k=\frac{E\big(g_k b_k
\big)^2}{E\big(\sum_{j=k+1}^{K}\rho_{kj}\hat{g}_j b_j\big)^2+N_{0}}.
\end {equation}

The main problem with MF and IC receivers in fading channels is the
unreliable estimation and hence cancellation of other users'
interference contributions, particularly when the users' channels
are time varying. It is well known that in CDMA, suppression of
interference leads to improved signal estimation of desired users,
leading to improved BER and system capacity performance. We can
assess the accuracy of the estimation performance of conventional
and blind adaptive receivers in terms of average MSE from all $K$
users, $\varepsilon^{2}$ given by
\begin{equation}
\label{eqn:min_mse}
\varepsilon^{2}=\frac{1}{K}\sum_{k=1}^{K}\varepsilon_{k}^{2}=\frac{1}{K}
\sum_{k=1}^{K}E\Big\{(\hat{g_k}\mid \hat{b}_k \mid- g_k \mid b_k
\mid)^{2}\Big\}.
\end{equation}

\section{Blind Adaptive IC Receivers}

Blind adaptive approaches to IC as in \cite{shakya_pic} and
\cite{shakya_sic2007} reduce the interference effects in two steps:

a) employ adaptive despreading to generate decision variables based
on minimum error cost function to preliminarily suppress the
interference,

b) use the signal estimates for interference cancellation that are
generated blindly from the despreader output and weighted by an
adaptive scaling factor at every symbol period $m$, $ 1 \leq m \leq
\infty$.

Unlike the conventional despreaders that multiply the received
signal with local copy of fixed amplitude spreading sequence, the
adaptive despreader uses weights $\textbf w_k(m)$ are updated every
symbol using the CMA algorithm employing a simple LMS type updating
\cite{haykin_adaptive_filter} with an objective to minimize the
error $e_{k}(m)= E\big\{\mid b_k(m)\mid
-\sum\textbf{w}_k(m)\big\}^2$ and is shown below
\begin{equation}
\textbf{w}_k(m+1)=\textbf w_k(m)-\mu\textbf{r}_k(m)e_k(m)
\end{equation}
where $\mu$ is the step size and $e_k$ is the instantaneous error of
CMA. Next, the scaling factor $\alpha_k(m)$ is obtained using the
despreader weights $\textbf{w}_k(m+1)$ and the spreading sequence
vector $\textbf{c}_k$ as follows
\begin{equation}
\label{eqn:est_scaling_factor1}
\alpha_k(m)=\frac{\breve{c}_k(m)}{\breve{w}_k(m)}
\end{equation}
where, $\breve{c}_k(m)$ and $\breve{w}_k(m)$ are the mean amplitude
of chips of user's spreading sequence and elements of the weight
vector, respectively and are given by
\begin{equation}
\breve{c}_k(m)=\frac{1}{N}\sum_{n=1}^{N} \Big|
c_k\big\{(m-1)N+n\big\}\Big| \end{equation} and
\begin{equation}\breve{w}_k(m)=\frac{1}{N}\sum_{n=1}^{N}
\Big|w_k\big\{(m-1)N+n\big\}\Big|.
\end{equation}

The signal ${z}_k(m)$ is then scaled with $\tilde {\alpha}_k(m)$ and
spread with $\textbf{c}_k$ to generate the cancellation term for
$k^{th}$ user for the next stage. Based on above principles, the
detection and estimation processes for BA-SIC and BA-PIC are given
next.

\noindent \textbf{BA-PIC} \cite{shakya_pic}: This scheme operates on
the same principles as the conventional PIC, however it obtains the
decision variable for the $k^{th}$ user at the $l^{th}$ PIC stage
$z^{BA-PIC(l)}_k(m)$ by utilizing the scaling factors as shown
above, obtained from the previous stage $\alpha^{l-1}_i(m)$ as
follows
\begin{equation}
\begin{split}
\label{eqn:est_scaling_factor1}
z^{BA-PIC(l)}_k(m)=\int_{0}^{Tb}\Big\{\textbf{r}(m)-\\
\sum_{j=1, j\neq
k}^{K}\alpha^{l-1}_j(m)z^{BA-PIC(l-1)}_j\textbf{c}_{j}\Big\}^{T}\textbf{w}_{k}(m);
\forall k.
\end{split}
\end{equation}
The user's channel estimate at the $l^{th}$ stage can be taken as
follows:
\begin{equation}
\hat{g_k}(m) b_k(m) = \alpha^{l}_k(m) z^{BA-PIC(l)}_k(m).
\end{equation}
The SINR of the BA-PIC at the $l$-th stage is obtained as follows:
\begin{equation}
\Gamma^{BA-PIC(l)}_k=\frac{E\big(g_k b_k
\big)^2}{var\big\{z^{BA-PIC(l-1)}_{k})\big\}}.
\end {equation}

Since the BA-PIC employs adaptive despreader at the initial stage,
the variance of interference at this stage is expected to lower than
conventional PIC. This is given in \cite{shakya_pic}, as follows:
\begin{equation}
var\{z^{BA-PIC(0)}_{k}\}=\frac{K}{M R( \Delta
t)}\sum_{m=1}^{M}\frac{\{e_{k}^{0}(m)\pm\mu\}^{2}}{N}.
\end {equation}
From above equation, we observe that the variance
$var\{z^{BA-PIC(0)}_{k}\}$ for a given time period depends upon
system parameters such as number of users in the system $K$, the
frame length $M$ considered ( usually $M>>1$), the channel fading
rate defined by the Doppler shift $R( \Delta t)$, the degree of
freedom for weight adaptation $N$ and the choice of step size $\mu$
\cite{haykin_adaptive_filter}.

Note that the choice of step size used in the adaptive algorithms
has direct effect on the detection performance. Although the step
size can also be varied adaptively to optimize the system
performance, this is left for future study. Therefore
$var\{z^{BA-PIC(0)}_{k}\}$ will be minimized, when $N$ is
sufficiently large and $R( \Delta t)$ does not change significantly
for the given frame period $M$. The following PIC stages improve the
detection performance significantly using adaptively weighted
interference cancellation using the despreader's weights leading
further reduced variance of interference. The variance at the $l$-th
stage of the BA-PIC is obtained as follows:
\begin{equation}
\begin{split}
var\{z^{BA-PIC(l)}_{j}\}=\varepsilon^{2}_k+\sum_{j=1,j \neq
           k}^{K}\varepsilon^{2}_j\Big\{\rho^{2}_{kj}-\kappa_{ji}\rho^{2}_{ki}\Big\}var\Big\{z^{l-1}_{ba-pic}\Big\};\\
           \kappa_{ji} = \left\{ \begin{array}{ll}
1 & \textrm{, if $  i=j        $}\\
-1 & \textrm{, if $   i\neq j  $}
\end{array} \right.;
           j=1,..,K; j\neq k .\\
\end{split}
\end {equation}

\noindent \textbf{BA-SIC} \cite{shakya_sic2007}: The decision
variable for the $k^{th}$ user, $z^{BA-SIC}_k(m)$ at each SIC stage
is obtained by selecting the despreader output of the strongest user
as follows:
\begin{equation}
\label{eqn:cmasic_dec} z^{BA-SIC}_k(m)=\max_{1 \leq u \leq
K}\Big\{{\int_{0}^{Tb}\textbf{r}_{k}^{T}(m)\textbf{w}_u(m)}\Big\}
\end{equation}
where $\textbf{w}_u(m)$ is the weight vector of the adaptive
despreader updated every symbol. The data and channel estimate of
the $k^{th}$ user can be obtained as follows:
\begin{equation}
\label{eqn:cmasic_chest} \hat{g_k}(m) b_k(m) =\alpha_k(m)
z^{BA-SIC}_k(m).
\end{equation}
The signal $z^{BA-SIC}_k$ is then respread using $\textbf{c}_k$ and
subtracted from remaining received signal to remove its interference
similar to that in (\ref{sic}). The SINR analysis for BA-SIC can be
obtained following similar approach for the BA-PIC as shown in
(17)-(19) and is not shown here for brevity.

It can be observed from the above equations that, the data detection
performance of all the receivers are dependent upon the MSE
performance of users' channel estimates. Hence we can conclude that
techniques that achieve lower MSE reduce interference and hence the
improved SINR or sum rate performances.
\section{Analysis of Achievable Sum Rates} In this
section, a simplified analysis of achievable sum rate or spectral
efficiency the system in a single isolated cell for the receivers is
provided assuming Gaussian distribution of interference \cite{Verdu}
and calculation of the expected SINR $E\{\Gamma\}$. Using
instantaneous SINR $\Gamma$ and distribution $p(\Gamma)$, the
multiple access capacity $C$ in a fading channel can be obtained as
follows:
\begin{equation}
\label{eqn:normal_se} C=\int_{0}^{\infty}
B\log_2\big[1+\Gamma]p(\Gamma)d\Gamma
\end{equation}
where $B$ is the signal bandwidth. In CDMA, the available bandwidth
is equally divided among $K$ users each occupying $1/N$ portion.
Therefore, the normalized spectral efficiency or sum rate $R_{sum}$,
is obtained as follows
\begin{equation}
\label{eqn:normal_se} R_{sum} \leq
\frac{K}{N}\log_2\big[1+E\{\overline{\Gamma}\}\big]
\end{equation}
where $\overline{\Gamma}=1/K\sum_{k=1}^{K}\Gamma_k$ is the average
SINR. Jensen's inequality theorem \cite{wireless_goldsmith} is
assumed here and hence the obtained spectral efficiency above
represents as the upper bound on what these techniques can actually
achieve in practice.
%The channel capacity in Rayleigh fading in Gaussian
%\noindent\emph{MF:} The spectral efficiency for conventional MF receivers can be
%obtained as $\Gamma^{MF} \leq
%{K}/{N}\log_2\big[1+E\{\overline{SINR}_{MF}\}\big]=
%{K}/{N}\log_2\big[1+{E^2\{z^{MF}\}\}}/{var\{z^{MF}\}}\big]$, where
%the expectation is taken over the fading power distribution and
%$var\{.\}$ is the variance.
%
%\noindent\emph{Conventional SIC:} For SIC the spectral efficiency can be obtained as
%follows: $\Gamma^{SIC} \leq
%{K}/{N}\log_2\big[1+E\{\overline{SINR}_{SIC}\}\big]=
%{K}/{N}\log_2\big[1+{E^2\{z^{SIC}\}}/{var\{z^{SIC}\}}\big]$.
%
%\noindent\emph{Conventional PIC:} The spectral efficiency of conventional PIC at
%$l^{th}$ stage is given by $ \Gamma^{PIC(l)}\leq
%{K}/{N}\log_2\big[1+E\{SINR^{PIC(l)}\}\big]=
%{K}/{N}\log_2\big[1+{E^2\{z^{PIC(l)}\}}/{var\{z^{PIC(l)}\}}\big]$.

\noindent \textbf{BA-PIC:} The spectral efficiency of BA-PIC at the
$l^{th}$ stage can be obtained as follows
\begin{equation}
\label{eqn:spectral_eff4}
\begin{split}
R_{sum}^{BA-PIC(l)}\leq
\frac{K}{N}\log_2\Big[1+E\big\{\overline{\Gamma}^{BA-PIC(l)}\big\}\Big]\\
=\frac{K}{N}\log_2\Bigg[1+\frac{1}{K}\sum_{k}\Gamma_{k}^{BA-PIC(l)}\Bigg]
\end{split}
\end{equation}
The lower variance of of BA-PIC compared with conventional PIC for
each stage is verified analytically and by simulations in
\cite{shakya_pic}. Therefore, it is expected to provide higher
spectral efficiency.

\noindent \textbf{BA-SIC:} The spectral efficiency of system
employing BA-SIC can be obtained as follows
\begin{equation}
\begin{split}
\label{eqn:se_cma_sic} R_{sum}^{BA-SIC} \leq
\frac{K}{N}\log_2\Big[1+E\big\{\overline{\Gamma}_{SIC}\big\}\Big]\\
=\frac{K}{N}\log_2\Bigg[1+\frac{1}{K}\sum_{k}\Gamma_{k}^{BA-SIC}\Bigg]
\end{split}
\end{equation}
Since $var\{z^{BA-SIC}\}$ will be much lower compared with
conventional SIC, it is anticipated that the sum rate will be higher
accordingly.

\section{Performance Study}
A baseband model of a synchronous uplink DS-CDMA with $K=20$ equal
power users i. e. average power control is assumed. Binary Gold
sequences of length $N = 31$ are used for spreading. The channel is
i.i.d Rayleigh flat fading for each user with normalized Doppler
rate $f_dT_b$ of 0.003 corresponding to mobile speed to $\approx
100$ km/hr at the carrier frequency of 2 GHz. A fixed step-size of
$\mu_k=0.0001$, is assumed in the adaptive despreader algorithm for
the detection of all users for the case of adaptive receivers. We
compare the system performance of conventional techniques, namely
the MF, PIC and SIC with the adaptive techniques in terms of MSE of
channel estimation and the spectral efficiency next.

\begin{figure}[tbp]
\begin{center}
\includegraphics[width=12.0 cm]{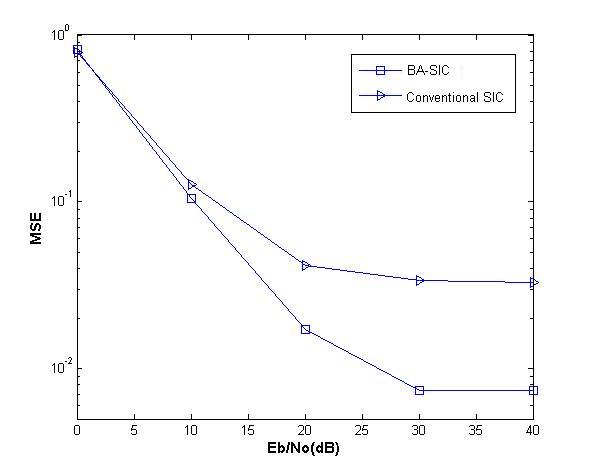}
\caption{MSE of channel estimation for the BA-SIC in Rayleigh flat
fading channel: $K=20$, $N=31$ (Gold sequence) and $f_d T_b=0.003$
are used.} \label{fig:sic_nonfading2}
\end{center}
\end{figure}

\begin{figure}[tbp]
\begin{center}
\includegraphics[width=12.0 cm]{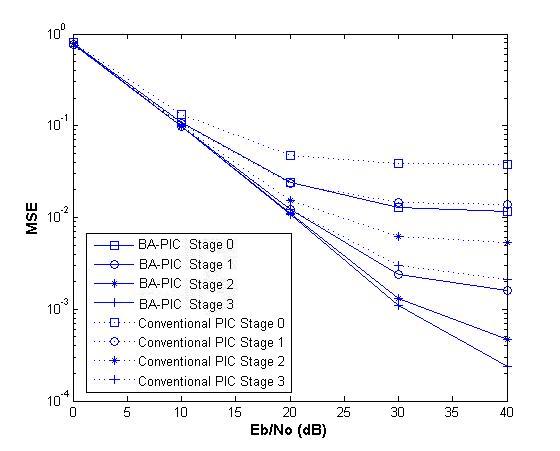}
\caption{MSE of channel estimation for the BA-PIC in Rayleigh flat
fading channel:$K=20$, $N=31$ (Gold sequence) and $f_d T_b=0.003$
are used.} \label{fig:pic_nonfading2}
\end{center}
\end{figure}

\noindent \textbf{MSE}: Figure \ref{fig:sic_nonfading2} shows the
MSE of channel estimation of conventional SIC and BA-SIC when users
employ binary PSK for data modulation. It can be clearly seen that
the adaptive technique offers much improved MSE as the $E_b/N_0$
increases. For example, an MSE of $0.0075$ compared with $0.034$ at
the $E_b/N_0=30$ dB is shown to be achieved. In Figure
\ref{fig:pic_nonfading2} the MSE performances of BA-PIC at different
interference cancellation stages are shown and compared to that with
conventional PIC. The BA-PIC shows much improved MSE for all stages.
For example, at the stage 1 it offers MSE of $0.002$ which is better
than that of stage 3 of conventional PIC with $0.0025$.

%The normalized spectral efficiency of conventional multistage PIC in
%flat Rayleigh fading channels is presented in Figure
%(\ref{fig_sim_spec_pic20}) for $K=20$ users. For the reference,
%performance under single user Rayleigh fading is also shown. As can
%be seen from the Figure, the use of interference cancellation
%gradually improves the achievable spectral efficiency of the
%receiver. This result is shown here only for comparison with
%proposed BA-PIC receiver.
%\begin{figure}[tbp]
%\centering
%\includegraphics[width=4.4 in]{CMA-PIC/Capacity_convPIC_stage0_3_20users_gold31_fading.jpg}
%% where an .eps filename suffix will be assumed under latex,
%% and a .pdf suffix will be assumed for pdflatex
%\caption{Spectral efficiency of Conventional PIC in Rayleigh flat
%fading channels, K=20, N=31 (Gold sequence)}
%\label{fig_sim_spec_pic20}
%\end{figure}

\begin{figure}[tbp]
\begin{center}
\includegraphics[width=12.0 cm]{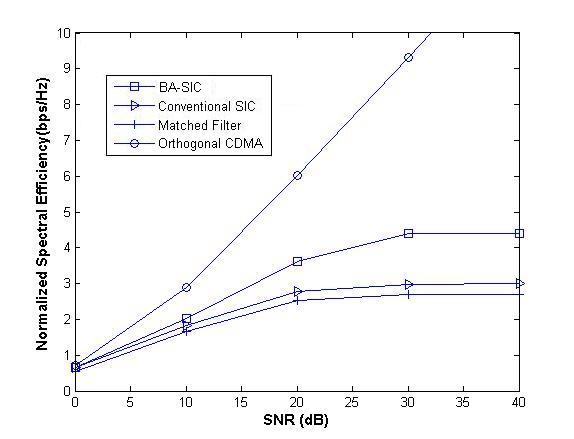}
\caption{Spectral Efficiency of BA-SIC in Rayleigh flat fading
channel, K=20, N=31 (Gold sequence)} \label{fig:sic_rates}
\end{center}
\end{figure}

\begin{figure}[tbp]
\begin{center}
\includegraphics[width=12.0 cm]{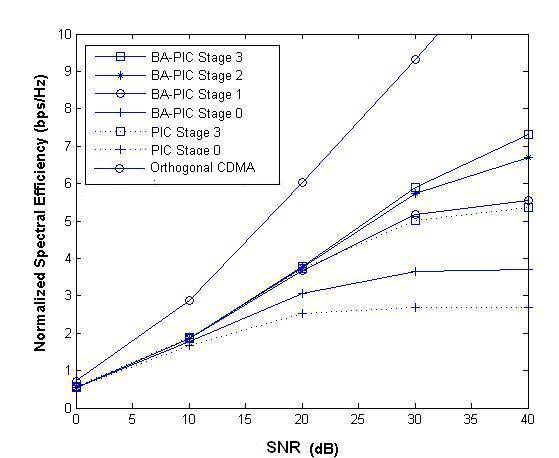}
\caption{Spectral Efficiency of BA-PIC in Rayleigh flat fading
channel, K=20, N=31 (Gold sequence)} \label{fig:pic_rates}
\end{center}
\end{figure}

\noindent \textbf{Spectral Efficiency}: Achievable spectral
efficiency of BA-SIC using the expression in (\ref{eqn:se_cma_sic})
is shown in Figure \ref{fig:sic_rates}. The spectral efficiency of a
fully loaded synchronous orthogonal CDMA which is equivalent to a
single user system \cite{spec_verdu}, is also shown for comparison.
As expected, the BA-SIC shows considerable gain compared with MF and
conventional SIC giving $\approx 4.4$ bits/s at the SNR of 30 dB
compared with $\approx 2.7$ and $\approx 3$ bits/s, respectively. It
can also be noted that, the conventional SIC offers only slightly
higher spectral efficiency compared with MF. This is due to
imperfect estimation and cancellation of the existing SIC methods
\cite{andrews_2005}. The spectral efficiency of BA-PIC using
(\ref{eqn:spectral_eff4}) with multiple stages of cancellation is
shown in Figure \ref{fig:pic_rates}. It can be clearly seen from the
figure that the scheme offers significant increase in the achievable
rate (spectral efficiency) compared with the conventional PIC. For
example, even with single stage of cancellation the BA-PIC shows the
performance comparable with the conventional PIC with three stage of
cancellation, providing $\approx 5$ bits/s at the SNR of 30 dB. From
the figures \ref{fig:sic_rates} and \ref{fig:pic_rates}, we can also
compare the sum rate or the spectral efficiency of BA-SIC and BA-PIC
under the same system conditions. It can be seen that the BA-PIC
offers much higher rate compared with BA-SIC as the number of
cancellation stages $l$ increases. For example, even with a single
stage of cancellation, BA-PIC gives the rate of $\approx 5.5$ bits/s
at the SNR of $30$ db, which is much higher than $\approx 4.5$
bits/s of the BA-SIC.

%\begin{figure}[tbp]
%\begin{center}
%\includegraphics[width=9.6 cm]{Capacity_CMAPIC_CMASIC_20users_gold31_fading.jpg}
%\caption{Spectral Efficiency of CMA-SIC in Rayleigh flat fading
%channel, K=20, N=31 (Gold sequence)} \label{fig:sic_nonfading2}
%\end{center}
%\end{figure}

\section{Conclusions}
We investigated the channel estimation performance of different
multiuser detection techniques and its impacts on the SINR and
achievable spectral efficiency in highly mobile Rayleigh fading
channel environments. Significantly improved MSE and sum data rates
are shown to be achieved with adaptive interference canceling
receivers. The adaptive receivers are shown to achieve much higher
sum rates, for example, $\approx 7.3$ bits/s compared with $\approx
5.3$ bits/s for the case of using parallel interference
cancellation, and also $\approx 4.5$ bits/s compared with $\approx
3.0$ bits/s for successive cancellation architecture, respectively.
For the future work, more efficient adaptive interference
suppression techniques for joint carrier frequency offset estimation
and cancellation for OFDMA will be carried out.

\bibliographystyle{unsrt}
%\bibliography{references}

% that's all folks

\end{document}